\documentclass[a4paper]{article}
\usepackage[]{graphicx}\usepackage[]{color}
\makeatletter
\def\maxwidth{ %
  \ifdim\Gin@nat@width>\linewidth
    \linewidth
  \else
    \Gin@nat@width
  \fi
}
\makeatother

\definecolor{fgcolor}{rgb}{0.345, 0.345, 0.345}

\usepackage{framed}
\makeatletter
\newenvironment{kframe}{%
 \def\at@end@of@kframe{}%
 \ifinner\ifhmode%
  \def\at@end@of@kframe{\end{minipage}}%
  \begin{minipage}{\columnwidth}%
 \fi\fi%
 \def\FrameCommand##1{\hskip\@totalleftmargin \hskip-\fboxsep
 \colorbox{shadecolor}{##1}\hskip-\fboxsep
     \hskip-\linewidth \hskip-\@totalleftmargin \hskip\columnwidth}%
 \MakeFramed {\advance\hsize-\width
   \@totalleftmargin\z@ \linewidth\hsize
   \@setminipage}}%
 {\par\unskip\endMakeFramed%
 \at@end@of@kframe}
\makeatother

\definecolor{shadecolor}{rgb}{.97, .97, .97}
\definecolor{messagecolor}{rgb}{0, 0, 0}
\definecolor{warningcolor}{rgb}{1, 0, 1}
\definecolor{errorcolor}{rgb}{1, 0, 0}
\newenvironment{knitrout}{}{} 

\usepackage{alltt} 
\usepackage{natbib} 
\usepackage[american]{babel}
\usepackage[utf8]{inputenc}
\usepackage{csquotes}

\usepackage{setspace}

\usepackage[outdir=./]{epstopdf}

\usepackage{amsmath,amssymb,amsfonts}

\usepackage{url}   
\usepackage{graphicx}   
\usepackage{verbatim}   
\usepackage{caption}
\usepackage{pdflscape}

\usepackage{fancyvrb}

\usepackage{newfloat}
\DeclareFloatingEnvironment[
]{listing}

\title{Statistical methods for linguistic research: Foundational Ideas - Part II}

\author{Bruno Nicenboim and Shravan Vasishth}

\begin{document}

\maketitle

\bibliographystyle{apalike}

\begin{abstract}
We provide an introductory review of Bayesian data analytical methods, with a focus on applications for linguistics, psychology, psycholinguistics, and cognitive science. The empirically oriented researcher will benefit from making Bayesian methods part of their statistical toolkit due to the many advantages of this framework, among them easier interpretation of results relative to research hypotheses, and flexible model specification. We present an informal introduction to the foundational ideas behind Bayesian data analysis, using, as an example, a linear mixed models analysis of data from a typical psycholinguistics experiment. We discuss hypothesis testing using the Bayes factor, and model selection using cross-validation. We close with some examples illustrating the flexibility of model specification in the Bayesian framework. Suggestions for further reading are also provided.
\end{abstract}

\section{Introduction}

In Part I of this review, we presented the main foundational ideas of frequentist
statistics, with a focus on their applications in linguistics and 
related areas in cognitive science. Our main goal there was to try to clarify the underlying
ideas in frequentist methodology. We discussed the 
meaning of the p-value, and of Type I and II errors  and Type S and M errors. 
We also discussed some common misinterpretations associated with hypothesis
testing in this framework.

There is no question that frequentist data-analytic methods such as the 
t-test, ANOVA, etc.\ are and will remain 
an important part of the toolkit of the experimentally inclined linguist. 
There is, however, another framework that is not yet part of the standard 
statistics curriculum in linguistics, but can be of great value: 
Bayesian data analysis. 
Learning to use this framework for data analysis is easier than it seems; 
in fact, most of us already think like Bayesians  when we carry out our frequentist  analyses.  One example of this is the (incorrect) interpretation of the p-value as the  probability of the null hypothesis being true. If we are faced with a p-value of 0.06 (i.e., a value just above the threshold of statistical significance), we often resort to expressions of gradedness, saying that ``the result was marginally significant'', or ``the results did not reach the conventional level of significance'', or even ``a non-significant trend towards significance.''\footnote{For a list of over 500 remarkably imaginative variants 
on this phrasing, see 
https://mchankins.wordpress.com/2013/04/21/still-not-significant-2/.}
This desire to take the non-significant result as meaningful comes about because 
of an irresistable temptation to (erroneously) give a Bayesian interpretation to
the p-value as the probability of the null
hypothesis being true. As discussed in Part I, the p-value is the probability
that the statistic is at least as extreme as the one observed given that
the null is true. When we misinterpret the p-value in terms of the 
probability of the null being true, 
a p-value of 0.06 doesn't seem much different from 0.04. 
Several studies have shown that such interpretations of the p-value are not 
uncommon (among many others: \citet{HallerKrauss2002,LecoutreEtAl2003}).
The interpretation of frequentist confidence intervals also suffers from 
similar problems  \citep{Hoekstra2014}. Given that the Bayesian interpretation 
is the more natural one that we converge to anyway, why not simply
do a Bayesian analysis? One reason that Bayesian methods have not become 
mainstream may be that, until recently,  
it was quite difficult to carry out a Bayesian analysis except in very limited 
situations. With the increase in computing power, and the arrival of 
several probabilistic programming languages, it has become quite easy to 
carry out relatively complicated analyses.

The objective of this paper is to provide a non-technical but practically 
 oriented review of some of the tools currently available for Bayesian data analysis. 
Since the linear mixed model (LMM) is so important for linguistics and related areas, 
we focus on this model and mention some extensions of LMMs. We assume here that the
reader has fitted frequentist linear mixed models  
\citep{lme4new}.
The review could also be used to achieve a better understanding of
papers that use Bayesian methods for statistical inference. 

We start the review by outlining what we consider to be the
main advantages of adopting Bayesian data-analytic methods. 
Then we informally outline the basic idea behind
Bayesian statistics: the use of Bayes' theorem to incorporate prior
information to our results. Next, we review several ways to verify the
plausibility of a research hypothesis with a simple example from 
psycholinguistics using Bayesian linear mixed models.
In the second part of the paper, we
discuss some example applications involving standard and less standard 
(but very useful) models. In the final section, we suggest some 
further readings that provide a more detailed presentation.

\section{Why bother to learn Bayesian data analysis?}

Statisticians have been arguing for decades about 
the relative merits of the frequentist vs Bayesian statistical methods.
We feel that both approaches, used as intended, have their merits, 
but that the importance of Bayesian approaches remains greatly 
underappreciated in linguistics and related disciplines. 

We see two main advantages to using Bayesian methods for data analysis. First,
Bayesian methods allow us to directly answer the question we are interested in:  
How plausible is our hypothesis given the data? 
We can answer this question by quantifying our uncertainty about the parameters 
 of interest.
Second, and perhaps more importantly, it is easier to flexibly define hierarchical 
models (also known as mixed effects or multilevel models) in the Bayesian
framework than in the frequentist framework.
Hierarchical models, whether frequentist or Bayesian, are highly relevant
for the repeated measures designs used in linguistics and psycholinguistics,
because they take both between- and within-group variances into
account, and because they pool information via ``shrinkage'' 
(see \citet{GelmanEtAl2012}). 
These properties have the desirable effects that 
we avoid overfitting the data, and we avoid averaging and
losing valuable information about group-level variability 
(\citet{gelmanhill07} provide more details). For example, both subjects and
items contribute independent sources of variance in a standard linguistics 
repeated measures design. In a hierarchical model, both these sources of 
variance can be included simultaneously. By contrast, in repeated measures 
ANOVA, one has to aggregate
by items (subjects), which artificially eliminates the variability between 
items (subjects). This aggregation leads to artificially 
small standard errors of the effect of interest, which leads to 
an increase in Type I error. 

The frequentist linear mixed model standardly used in psycholinguistics
is generally
fit with the \texttt{lme4} package \citep{lme4new} in R. However, if we want to
include the maximal random effects structure justified by the design
\citep{SchielzethForstmeier2009,barr2013random}, these models tend to not
converge or to give unrealistic estimates of the correlations between random 
effects \citep{BatesEtAlParsimonious}. In contrast, the maximal random effects
structure can be fit without problems using Bayesian methods, as discussed later in this review 
\citep[also see][]{chung2013weakly,BatesEtAlParsimonious,SorensenVasishthTutorial}. 
In addition, Bayesian
methods can allow us to hierarchically extend  virtually any model: non-linear
models (which are not generalized linear models) and even the highly complex
models of cognitive processes \citep{Lee2011,ShiffrinEtAl2008}. See
also the discussion about hierarchical models in the 
section entitled
Examples of applications of Bayesian methods.

\section{Bayesian data analysis: An informal introduction}

In linguistics, we are usually interested in determining
whether there is an effect of a particular factor on
some dependent variable; an example from psycholinguistics 
is the difference in 
processing difficulty between subject and object relative clauses as measured
by reading times. In the frequentist paradigm, we assume that 
there is some unknown point value $\mu$ that represents the difference
in reading time between the two relative clause types; the goal is to reject
the null hypothesis that this true $\mu$ has the value $0$. 
In the Bayesian framework,
our goal is to obtain
an estimate of $\mu$ given the data along
with an uncertainty estimate (such as a credible interval, discussed
in detail below) that
gives us a range over which we can be reasonably sure that the true
parameter value lies. We obtain these estimates given the data and
given our prior knowledge/information about plausible values of
$\mu$.  We elaborate on this idea in the next section when we present
a practical example, but the essential point is that the 
distribution of $\mu$ (called the posterior distribution)
can be expressed in terms of the prior and 
likelihood:\footnote{The term likelihood may be unfamiliar. For example, if we have
$n$ independent data points, $x_1, \dots, x_n$ which are assumed
to be generated from a Normal distribution with parameters $\mu$ and
$\sigma$ and a probability density
function $f(\cdot)$, the joint probability
of these data points is the product $f(x_1)\times f(x_2)\times \dots \times 
f(x_n)$. 
The value of this product is a function of different values of 
$\mu$ and $\sigma$, and it is common to call this product 
the Likelihood function, and it is often written 
$L(x_1,\dots,x_n; \mu,\sigma^2)$. The essential idea here is that  the 
likelihood tells us the joint probability of the data for different
values of the parameters.}

\begin{equation}
\hbox{Posterior} \propto \hbox{Prior} \times \hbox{Likelihood}
\end{equation}

\begin{figure}[!htbp]
\begin{knitrout}
\definecolor{shadecolor}{rgb}{0.969, 0.969, 0.969}\color{fgcolor}

{\centering \includegraphics[width=\maxwidth]{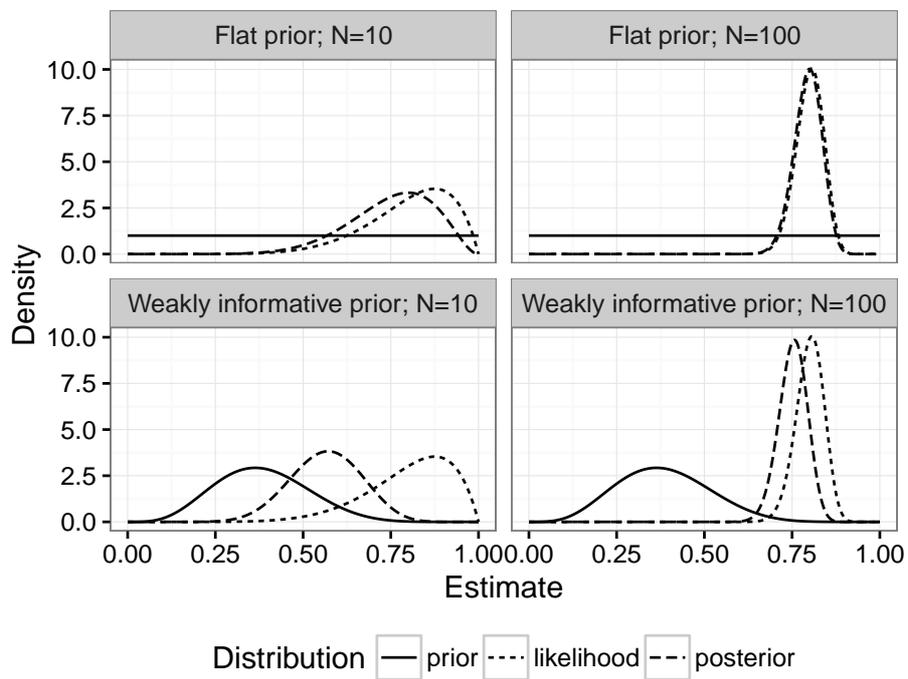} 

}

\end{knitrout}

\caption{Posterior distributions given different likelihoods and priors for binomial data.}\label{fig:difflikepriors}
\end{figure}

To repeat, given some prior information
about the parameter of interest, and the likelihood, 
we can compute the posterior distribution of the parameter. 
The focus is not on rejecting a null hypothesis 
but on  what the posterior distribution tells
us about the plausible values of the parameter of interest.
Recall that frequentist significance testing is focused on calculating 
the p-value $P( statistic \mid \mu=0)$, that is, the probability of 
observing a test statistic (such as a t-value) at least as extreme as
the one we observed given 
that the null hypothesis that $\mu=0$ is true. 
By contrast, Bayesian statistics allows us to talk about
plausible values of the parameter $\mu$ given the data, 
through the posterior distribution
of the parameter. 
Another important
point is that the posterior is essentially a weighted mean of 
the prior and the likelihood.   
The implications of this statement are made clear graphically in 
Figure
\ref{fig:difflikepriors}. Here, we see the effect of two types of prior 
distributions on
binomial data given 10 or 100 observations. 
Two points are worth noting. First,
when the prior is spread out over a wide range and assign equal probability to all possible values, the posterior 
distribution ends up closer to the likelihood; this alignment to the 
likelihood is more pronounced when we have more data (larger sample size).
Second, when we have weakly informative priors, with sparse data, 
the posterior is closer to the prior, but with more data, 
the posterior is again closer to the likelihood. 
What this implies is that when we have little data, it is worth
investing time in developing priors informed by prior knowledge; 
but when we have a lot of data, the likelihood will dominate in determining
the posterior (we return to this point later, with an example). 
Indeed, in large-sample situations, we will usually find that 
the Bayesian posterior and the frequentist mean, along with their uncertainty 
estimates, are nearly identical or very similar (even though their meaning is
quite different---see the discussion below on Bayesian credible intervals).

Bayes' theorem is just a mathematical rule that allows
us to calculate any posterior distribution. In practice, however, this is true
for a very limited number of cases, and, in fact, the posterior of many of the
models that we are interested in  cannot be derived  analytically. Fortunately, the
posterior distribution can be approximated with numerical techniques such as
Markov Chain Monte Carlo (MCMC). Many of the 
probabilistic programming languages
freely available today (see the final section for a listing) 
allow us define our models without having to acquire expert knowledge
about the relevant numerical techniques. 

It is all very well to talk about Bayesian methods in the abstract, 
but how can they be used by linguists and psycholinguists? 
To answer this question, consider a concrete example from 
psycholinguistics. We feel that it is easier to show an example that the
reader can relate to in order to convey a feeling for how a Bayesian analysis
would work; for further study, excellent introductory textbooks are available
(we give some suggestions in the final section).

\section{An example of statistical inference using Bayesian methods}

In contrast to significance testing in frequentist
statistics, Bayesian inference 
is not necessarily about dichotomous decisions (reject null or 
fail to reject null), 
but rather about the evidence for and against
different hypotheses.  
We illustrate one way to carry out statistical inference
using a Bayesian linear mixed model with the data from \citet{gibsonwu}, which
has a simple two-condition repeated measures design.
\citet{gibsonwu} compared reading times at the head noun for subject and
object relative clauses and argued for facilitation in the case of object
relative clauses (providing a counter-example to the cross-linguistic
generalization that
subject relative clauses are easier to process 
than object relative clauses; also see \citet{hsiao03}). 
We will assume, as we have done elsewhere \citep{SorensenVasishthTutorial},
that the dependent variable,
reading times, has a lognormal distribution, and thus we will use 
log reading times as the dependent variable.\footnote{Notice, however, that
this is not necessarily the best characterization of latencies; see
\citet{Ratcliff1993,Rouder2005,NicenboimEtAlFrontiers2015Capacity}.} We
fit the linear mixed model with the full random effects
structure justified by the design \citep{barr2013random}, 
and we code object
relative clauses with $1$ and subject relative clauses with $-1$. With this sum contrast coding,
\citeauthor{gibsonwu}'s prediction that object relative clauses are easier
than subject relative clauses in Chinese would, if correct, result in an  
effect with a negative sign (i.e. shorter reading times for the condition coded as 1).
In the Gibson and Wu dataset, we have 37 participants and 15 items; due to some 
missing data, we have a total of 547 data points.
The  conditions were presented to participants in a counterbalanced manner 
using a Latin square. 

The researcher familiar with \texttt{lme4}
(for details see
\citet{batessarkar} and \texttt{lme4} documentation: \citet{lme4new})
will not find the transition to the Bayesian approach difficult. 
In \texttt{lme4} syntax, the model we would fit would be

\begin{verbatim}
lmer(log(rt) ~ cond + (cond|subj)+(cond|item))
\end{verbatim}

We can fit an analogous Bayesian linear mixed model with the 
\texttt{stan\_lmer} function from the \texttt{rstanarm} package 
\citep{rstanarm2016}; see the code in Listing \ref{code:modelfitting}. 
The main novelty in the syntax is the specification of the priors for each 
parameter.
Some other details need to be specified, such as 
the desired number of chains and iterations that the MCMC
algorithm requires to converge to the posterior distribution of
the parameter of interest. To speed up computation, 
the number of processors (cores) available in their computer can also be 
specified. 
In the example shown in Listing~\ref{code:modelfitting}, 
we have left other parameters of the MCMC algorithm  at the default
values, but they may need some fine tuning in case of non-convergence
(this is announced by a warning message). 

A comparison of the estimates from the 
\texttt{lme4} ``maximal'' LMM and the analogous Bayesian
LMM are shown in Table~\ref{tab:lmerstancomparison}. An important point
to notice is that correlations between the varying intercepts and slopes
in the fitted model using the \texttt{lmer} function are on the boundary;
although this does not register a convergence failure warning in \texttt{lmer},
such boundary values constitute a failure to estimate the correlations 
\citep{BatesEtAlParsimonious}. 
This failure to estimate the correlations is due to the sparsity of data:
we have only 37 subjects and even fewer items (15) and are asking too much 
of the \texttt{lmer} function. In the Bayesian LMM,
the prior specified on the correlation matrix of the random effects ensures
that if there is insufficient data, the posterior will have a mean correlation 
near 0 with a wide uncertainty associated with it (this is the case in the
items intercept-slope correlation), and if there is more data, the posterior 
will be a compromise between the prior and the likelihood, although 
the uncertainty associated with this estimate may still be high 
(this is the case in the subjects intercept-slope correlation). See 
\citet{SorensenVasishthTutorial} for more discussion on this point.

\singlespacing
\begin{listing}
\begin{knitrout}
\definecolor{shadecolor}{rgb}{1, 1, 1}\color{fgcolor}\begin{kframe}
\begin{Verbatim}[numbers=left,fontfamily=courier,fontsize=\footnotesize, firstnumber=last]
library(rstanarm)
 \end{Verbatim}
\begin{Verbatim}[numbers=left,fontfamily=courier,fontsize=\footnotesize, firstnumber=last]
dgw <- read.table("gibsonwu2012data.txt")
 \end{Verbatim}
\begin{Verbatim}[numbers=left,fontfamily=courier,fontsize=\footnotesize, firstnumber=last]
dgw_hn <- subset(dgw, subset = region == "headnoun")
 \end{Verbatim}
\begin{Verbatim}[numbers=left,fontfamily=courier,fontsize=\footnotesize, firstnumber=last]
dgw_hn$cond <- ifelse(dgw_hn$type == "obj-ext", 1, -1)
 \end{Verbatim}
\begin{Verbatim}[numbers=left,fontfamily=courier,fontsize=\footnotesize, firstnumber=last]
m1 <- stan_lmer(formula = log(rt) ~ cond + (cond | subj) + (cond | item), 
                prior_intercept = normal(0, 10), 
                prior = normal(0, 1),            
                prior_covariance = decov(regularization = 2), 
                data = dgw_hn, 
                chains = 4, 
                iter = 2000, 
                cores = 4)
 \end{Verbatim}
\begin{Verbatim}[numbers=left,fontfamily=courier,fontsize=\footnotesize, firstnumber=last]
#summary(m1) # Very long summary with all the parameters in the model
 \end{Verbatim}
\end{kframe}
\end{knitrout}

\caption{Code for fitting a linear mixed model with stan\_lmer. A major difference from lme4 syntax is that priors are specified for (a) the intercept, (b) the slope, and (c) the variance-covariance matrices the random effects for subject and item. The other specifications, regarding
chains and iterations, relate to the how samples are taken from the posterior distribution, and specifying the number of cores can speed up computation.}\label{code:modelfitting}
\end{listing}

\begin{table}[!htbp]
\caption{Comparison of the frequentist and Bayesian model estimates for the 
Gibson and Wu data-set.  
The estimates for both the coefficients and the variance components 
are comparable, but notice that the correlations between varying
intercepts and slopes are quite different in the frequentist and
Bayesian models. The stan\_lmer packages provides
the median and the standard deviation of the median absolute 
difference (MAD) for the fixed effects, 
but one could equally well compute the mean and 
standard error, to mirror the lme4 convention.}
\begin{center}
\begin{tabular}{cccc|cc}
\multicolumn{4}{c}{\textbf{lmer}} & \multicolumn{2}{c}{\textbf{stan\_lmer}}\\
\hline
\multicolumn{4}{c}{\underline{Random effects}} & 
\multicolumn{2}{c}{\underline{Random effects}}\\
\hline
 Groups &  Name     &   Std.Dev. &  Corr  &  Std.Dev. & Corr\\  
 \hline
 subj   &  (Intercept)& 0.2448 &       & 0.2425      & \\
        &  cond       & 0.0595 &-1.00  & 0.0762 & -0.521\\ 
 item   &  (Intercept)& 0.1820 &      &  0.1829 &  \\     
        &  cond       & 0.0002 &1.00   & 0.0475 & 0.012\\ 
Residual &            & 0.5143 &     &   0.5131 &  \\
\hline
\multicolumn{4}{c}{\underline{Fixed effects}} & 
\multicolumn{2}{c}{\underline{Fixed effects}}\\
\hline
            &  Estimate &Std.\ Error& t value & Median &  MAD-SD\\
\hline            
(Intercept) & 6.06180   & 0.0657 &  92.24 & 6.0641 & 0.0658\\
cond        & -0.03625  &  0.0242  & -1.50 &-0.0364  & 0.0301\\
\hline
\end{tabular}
\end{center}
\label{tab:lmerstancomparison}
\end{table}%

\section{Prior specification} 
Since priors are an important part of the model, it is worth
saying a few 
words about them. In order to fit a Bayesian model, we need to 
specify a
prior distribution on each parameter; these priors express our initial 
state of knowledge about the
possible values that the parameter can have. 
It is possible to specify completely
\emph{uninformative priors}, such as flat priors, but these
 are far from ideal since they concentrate too much probability mass
outside of any reasonable posterior values. This  
can have the consequence that without enough
data, this prior will dominate in determining the posterior mean
and the uncertainty associated with it \citep{gelman2006prior}. 
Priors that give some minimal amount of
information improve inference and are called \emph{regularizing} or
\emph{weakly informative priors} (see also
\citet{chung2013weakly,gelman2008weakly}). 

In the typical psycholinguistic experiment, 
different weakly
informative priors generally don't have much of an effect on 
the posterior, but it is a good idea to do a \textit{sensitivity 
analysis} by evaluating the effect of different priors on the posterior;
see \citet{VasishthetalPLoSOne2013,Levshina2016} for examples from linguistics. 
Another use of priors is to include valuable prior knowledge about the parameters 
into our model; these \emph{informative priors}
could come from actual
prior experiments (the posterior from previous experiments) 
or from meta-analyses \citep{VasishthetalPLoSOne2013}, or from
expert judgements \citep{ohagan2006uncertain,Vasishth:MScStatistics}. 
Such a use of priors is not widespread even among Bayesian statisticians, but could be a powerful tool for incorporating prior knowledge into 
a new data analysis. For example, starting with relatively
informative priors could be a huge improvement over
starting every new study on relative clauses with the assumption 
that we know nothing about the topic. 

Returning to the practical issue of prior specification in R functions like, \texttt{stan\_lmer}, it is a good idea to specify the priors 
explicitly; the default priors assumed by the function may not be 
appropriate for your specific data.
For example, in our example above, if we set the prior for the intercept
as a normal distribution with a mean of zero and a standard deviation of ten,
Normal(0,10),\footnote{This happens to be the default value in this version of
rstanarm, but this might change in future versions.} we are assuming that we
are 68\% sure that the grand mean in log-scale will be between $-10$ (very
near zero milliseconds) and $10$ ($\approx 22$ seconds). This prior is
extremely vague, and can be changed but it won't affect the results much.
However, notice that had we not log-transformed the dependent variable, we
would be assuming that we are 68\% sure that the grand mean is between -10 and 10
milliseconds, and 95\% sure that it is between -20 and 20 milliseconds! 
We are guaranteed to not get sensible estimates if we do that.

We start the analysis by setting  Normal(0,1) as the prior for the effect of
subject vs.\ object relative clauses. This means that we assume that we are
68\% certain that the difference between the conditions should be less than
1300 ms.\footnote{The mean reading time at the critical region is 550ms, and
since we assumed a lognormal distribution, to calculate the ms, we need to
find out $\exp(\log(550)+1) - \exp(\log(550)-1)$, which is 1300; the 68\% is
the probability mass between [-1,1] in the standard normal distribution.} 
In reality, the range is likely to be much smaller, but we will start 
with this prior. We will exemplify the effect of the priors by changing the
prior of the estimate for the effect of the experimental condition in the next
section. The reader may notice that there is also a prior for the covariance
matrix of the random effects \citep{chung2013weakly}; the
regularization parameter in this prior specification 
to a value larger than one can help to get conservative estimates of the
intercept-slope correlations when we don't have enough data; recall
the discussion regarding Table~\ref{tab:lmerstancomparison} above. 
For further examples, see
\citet{BatesEtAlParsimonious} and the vignettes in the 
\texttt{RePsychLing} package  
(https://github.com/dmbates/RePsychLing).
 A more in-depth discussion is beyond the scope of this paper, and the 
 reader is referred to
the tutorial by \citet{SorensenVasishthTutorial} and a more general
discussion by
\citet{chung2013weakly}.\footnote{We left the priors of other parameters such
as the standard deviation of the distribution, or residuals in lme4 terms, and
the scale of the random effects at their default values. These priors
shouldn't be ignored when doing a real analysis! The user should verify that
they make sense.}

\section{The posterior and statistical inference}

As mentioned earlier, the result of a Bayesian analysis is a posterior
distribution, that is, a distribution showing the relative plausibilities 
of each possible value of
the parameter of interest, conditional on the data, the priors, and the model.
Every parameter of the model (fixed effects, random effects, shape of the
distribution) will have a posterior distribution. Typically, software 
such as the \texttt{rstanarm} package in R
will deliver samples from the posterior that we can use for 
inference. In our running example, we will focus on the
posterior distribution of the effect of object relative clauses in comparison
with subject relative clauses; see Figure \ref{fig:posterior}. 

To communicate our results, we need  to summarize and interpret the posterior
distribution. We can report point estimates of  the posterior probability such
as  the mean or the median (in some cases also the mode of the distribution,
known as the \emph{maximum a posteriori} or MAP, is also reported). When the
posterior distribution is symmetrical and approximately normal in shape, 
the mean and median
almost converge to the same point (see Figure \ref{fig:posterior}). In this
case it won't matter which one we report,  since  they only differ  in the
fourth decimal digit; see Listing \ref{code:medianmean}.

\singlespacing
\begin{listing}
\begin{knitrout}
\definecolor{shadecolor}{rgb}{1, 1, 1}\color{fgcolor}\begin{kframe}
\begin{Verbatim}[numbers=left,fontfamily=courier,fontsize=\footnotesize, firstnumber=last]
samples_m1 <- as.data.frame(m1) # It saves all the samples from the model.
 \end{Verbatim}
\begin{Verbatim}[numbers=left,fontfamily=courier,fontsize=\footnotesize, firstnumber=last]
posterior_condition <- samples_m1$cond
 \end{Verbatim}
\begin{Verbatim}[numbers=left,fontfamily=courier,fontsize=\footnotesize, firstnumber=last]
options(digits = 4) 
 \end{Verbatim}
\begin{Verbatim}[numbers=left,fontfamily=courier,fontsize=\footnotesize, firstnumber=last]
mean(posterior_condition) 
 \end{Verbatim}
\begin{Verbatim}[fontfamily=courier,fontsize=\footnotesize, firstnumber=last]
## [1] -0.0358
 \end{Verbatim}
\begin{Verbatim}[numbers=left,fontfamily=courier,fontsize=\footnotesize, firstnumber=last]
median(posterior_condition)  
 \end{Verbatim}
\begin{Verbatim}[fontfamily=courier,fontsize=\footnotesize, firstnumber=last]
## [1] -0.03573
 \end{Verbatim}
\end{kframe}
\end{knitrout}
\caption{Code for summarizing point estimates.}\label{code:medianmean}
\end{listing}

It is also important to summarize  the amount of posterior probability that
lies below or above some parameter value. 
In \citeauthor{gibsonwu}'s data, since the research question amounts to 
whether the parameter is positive or negative, 
and Gibson and Wu predict that it will be negative,
we can compute the posterior 
probability that
the difference between object and subject relative clauses is less than zero,
$P(\hat{\beta})<0$. This probability is 0.89;  
see Listing \ref{code:below0} and
also Figure \ref{fig:posteriorbelows}(a).

\singlespacing
\begin{listing}
\begin{knitrout}
\definecolor{shadecolor}{rgb}{1, 1, 1}\color{fgcolor}\begin{kframe}
\begin{Verbatim}[numbers=left,fontfamily=courier,fontsize=\footnotesize, firstnumber=last]
mean(posterior_condition < 0) # Posterior probability that lies below zero. 
 \end{Verbatim}
\begin{Verbatim}[fontfamily=courier,fontsize=\footnotesize, firstnumber=last]
## [1] 0.89
 \end{Verbatim}
\end{kframe}
\end{knitrout}
\caption{Code for finding the mass of the posterior probability that lies
below zero.}\label{code:below0}
\end{listing}

There is nothing special about zero; 
and since the difference between English object and subject
relative clauses is, in general, quite large, and
\citeauthor{gibsonwu} predict that in Chinese 
the same processes as in English give an
advantage to object relatives over subject relative clauses, we could be
interested, instead, in knowing the probability that the advantage of object
relative clauses is at least 20 ms. This advantage can be
translated into approximately $\ensuremath{-0.02}$ from the grand mean in
log-scale; we can inspect the posterior distribution and find out that this,
$P(\hat{\beta})<\ensuremath{-0.02}$, is 0.67; see also
Figure \ref{fig:posteriorbelows}(b).

\begin{figure}[!htbp]
\begin{knitrout}
\definecolor{shadecolor}{rgb}{0.969, 0.969, 0.969}\color{fgcolor}

{\centering \includegraphics[width=\maxwidth]{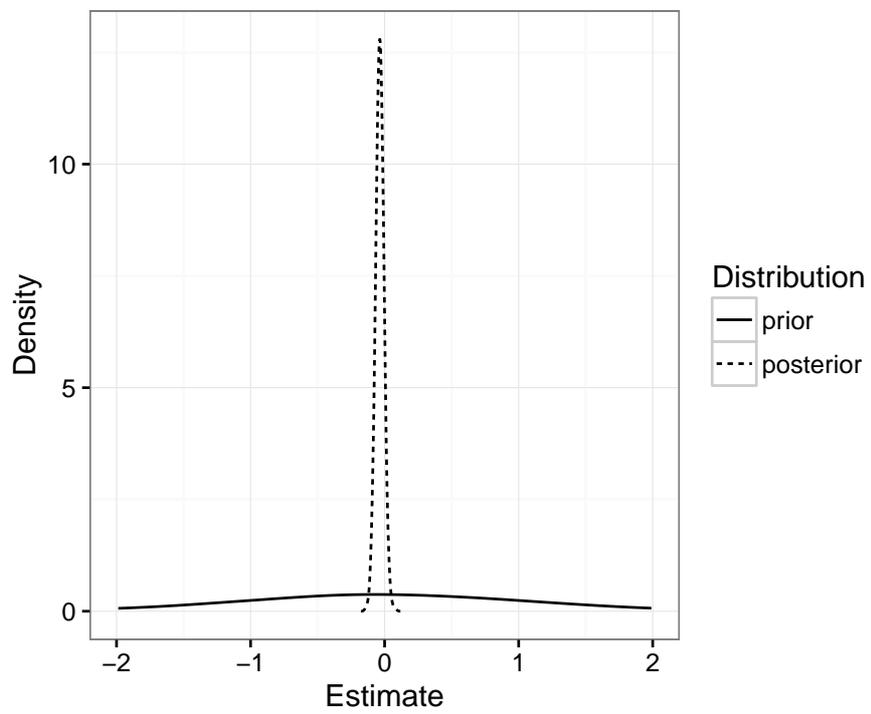} 

}

\end{knitrout}
\caption{Posterior distribution of the difference between object and subject
relative clauses given a prior distribution Normal(0,1).}
\label{fig:posterior}
\end{figure}

\begin{figure}[!htbp]
\begin{knitrout}
\definecolor{shadecolor}{rgb}{0.969, 0.969, 0.969}\color{fgcolor}

{\centering \includegraphics[width=\maxwidth]{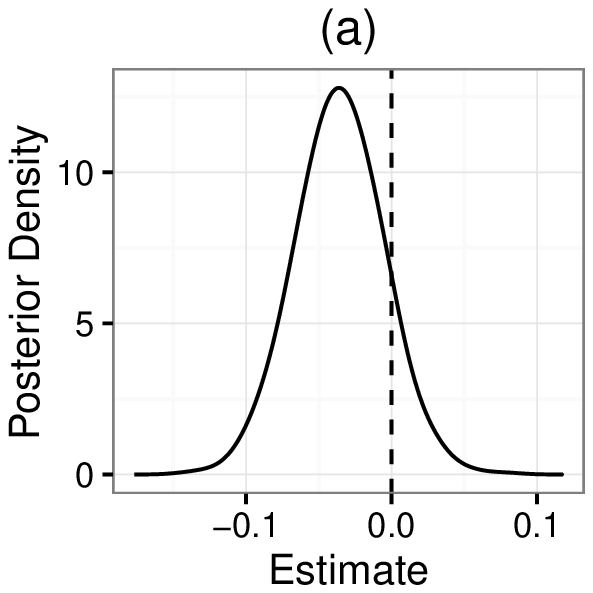} 
\includegraphics[width=\maxwidth]{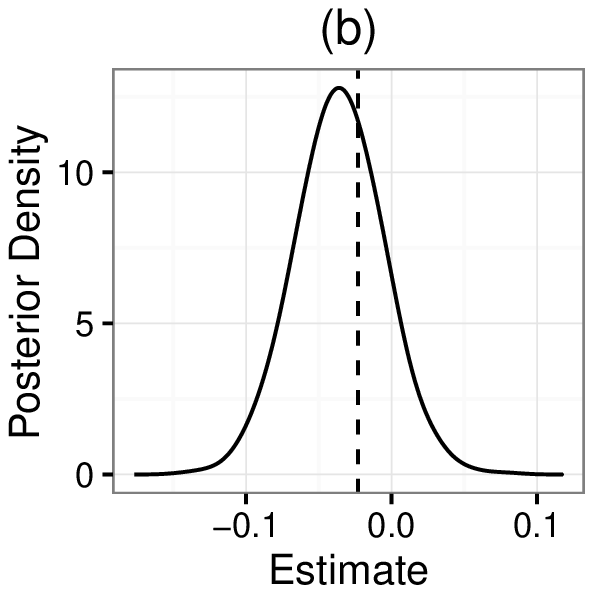} 

}

\end{knitrout}
\caption{The posterior probability that the difference between object and subject relative clauses is less than zero (a), and less than \ensuremath{-0.02} (b).}
\label{fig:posteriorbelows}
\end{figure}

\subsection{The 95\% credible interval}

It is possible (and desirable) to report an interval of posterior probability,
that is, two parameter values that contain between them a specified amount of
posterior probability. This type of interval is also known as \emph{credible
interval}. A credible interval demarcates the range within which we can be
certain with a certain probability that the ``true value'' of a parameter
lies. It is the true value not out there ``in nature'', but true in the
model's logical world (see also the interesting distinction between
\emph{small} and \emph{large worlds} in \citet{mcelreath2015statistical}).

The Bayesian credible interval is different from the frequentist confidence
interval because the credible interval can be interpreted with the data at
hand, while the frequentist counterpart is a property of the statistical
procedure. The statistical procedure only indicates that frequentist confidence intervals
across a series of hypothetical data sets produced by the same underlying
process will contain the true parameter value in a certain proportion of the
cases \citep{Hoekstra2014,MoreyEtAl2015}.

 The two most common types of Bayesian credible intervals are the percentile
interval and highest posterior density interval
\citep[HPDI;][]{box2011bayesian}. In the first one, we assign equal
probability mass to each tail. This is the most common way to report credible
intervals, because non-Bayesian intervals are usually percentile intervals. As
with frequentist confidence intervals, it is common to report 95\%  intervals
(see Figure \ref{fig:CI}). The second option is to report the HDPI, that  is
the narrowest interval containing the specified probability mass. This
interval will show the parameter values most consistent with the data, but it
can be noisy and depends on the sampling process
\citep{liu2013simulation}. When the posterior is symmetrical and normal
looking, it is very similar to the percentile interval: In the case of
\citeauthor{gibsonwu}'s data, the difference between them is in the second or third
decimal digit; see Listing \ref{code:CrI}.

\singlespacing
\begin{listing}
\begin{knitrout}
\definecolor{shadecolor}{rgb}{1, 1, 1}\color{fgcolor}\begin{kframe}
\begin{Verbatim}[numbers=left,fontfamily=courier,fontsize=\footnotesize, firstnumber=last]
options(digits = 4) 
 \end{Verbatim}
\begin{Verbatim}[numbers=left,fontfamily=courier,fontsize=\footnotesize, firstnumber=last]
posterior_interval(m1, par = "cond", prob = 0.95) # 95% Percentile Interval
 \end{Verbatim}
\begin{Verbatim}[fontfamily=courier,fontsize=\footnotesize, firstnumber=last]
##          2.5%   97.5%
## cond -0.09553 0.02427
 \end{Verbatim}
\begin{Verbatim}[numbers=left,fontfamily=courier,fontsize=\footnotesize, firstnumber=last]
library(SPIn) # For calculating the HPDI
 \end{Verbatim}
\begin{Verbatim}[numbers=left,fontfamily=courier,fontsize=\footnotesize, firstnumber=last]
bootSPIn(posterior_condition)$spin # 95% HPDI  
 \end{Verbatim}
\begin{Verbatim}[fontfamily=courier,fontsize=\footnotesize, firstnumber=last]
## [1] -0.09563  0.02355
 \end{Verbatim}
\end{kframe}
\end{knitrout}
\caption{Code for 95\% Percentile Interval and HPDI.}\label{code:CrI}
\end{listing}

\begin{figure}[!htbp]
\begin{knitrout}
\definecolor{shadecolor}{rgb}{0.969, 0.969, 0.969}\color{fgcolor}

{\centering \includegraphics[width=\maxwidth]{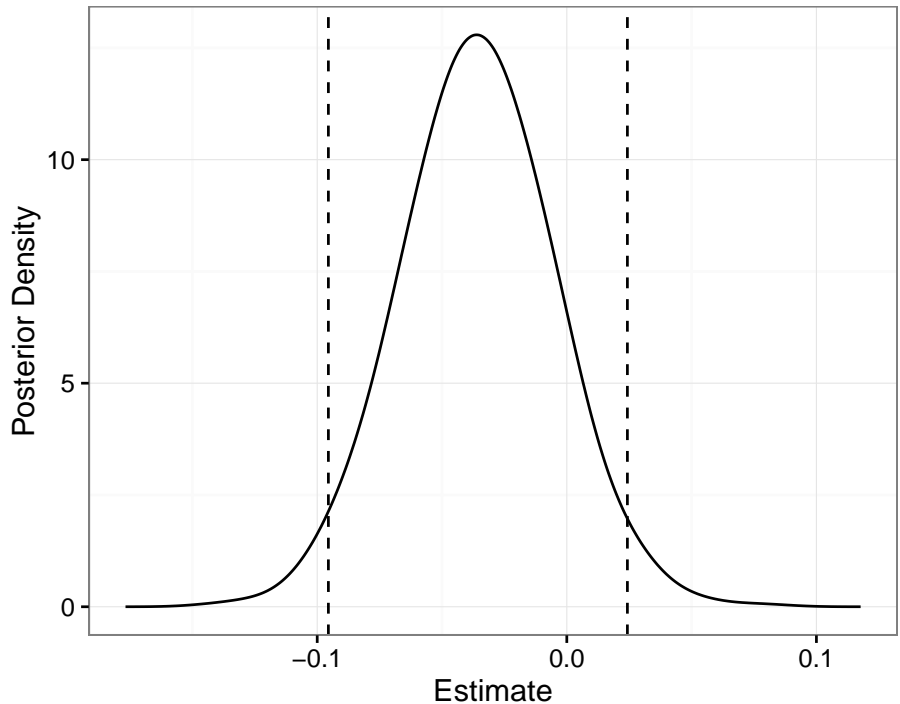} 

}

\end{knitrout}
\caption{95\% Credible Interval for the effect in the Gibson and Wu analysis.}
\label{fig:CI}
\end{figure}

\subsection{Investigating the effect of prior specification on posteriors}
\begin{figure}[!htbp]
\begin{knitrout}
\definecolor{shadecolor}{rgb}{0.969, 0.969, 0.969}\color{fgcolor}

{\centering \includegraphics[width=\maxwidth]{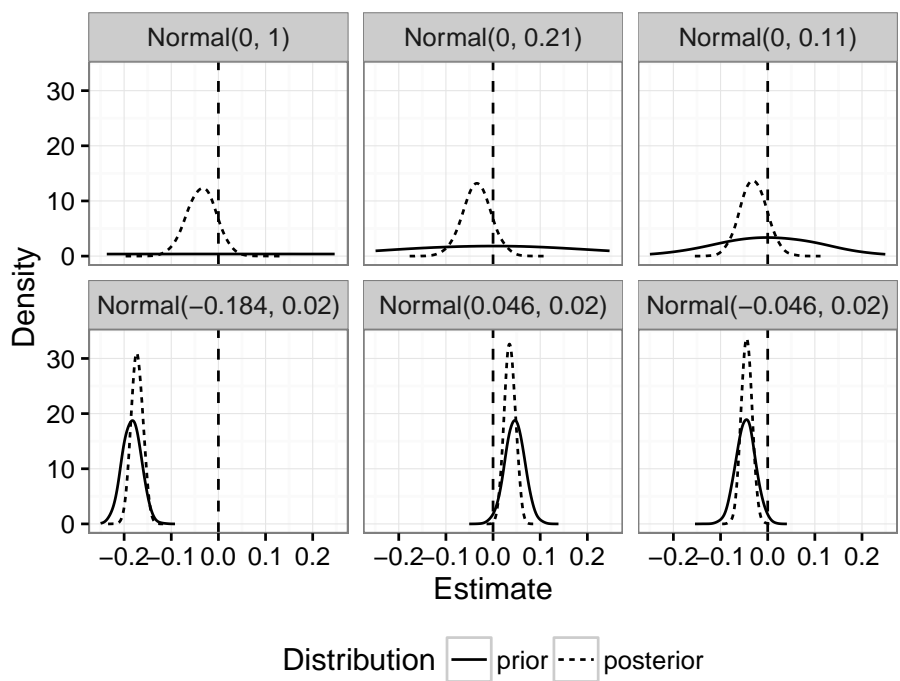} 

}

\end{knitrout}

\caption{Posterior distributions given different type of priors. The first row
shows different weakly informative priors, while the second row
shows unreasonably constrained priors.}\label{fig:diffpriors}
\end{figure}

So far we dealt with a very weakly informative prior; what would happen with
more informative ones? Let us start by choosing a reasonable alternative prior, such as
$Normal(0,0.21)$ or
$Normal(0,0.11)$; these assume that the difference
between conditions will be around 200 or
100 ms, and can be positive or negative; see the top row of Figure
\ref{fig:diffpriors}. 
Alternatively, we could have chosen unrealistically tightly constrained 
priors, such as (a)
$Normal(0.02,
0.02)$,
which assumes a difference four time larger between subject and object
relative clauses than what the data shows; (b)
$Normal(0.05,
0.02)$, which
assumes  that object relative clauses \emph{are slower} than subject
relative clauses also in Chinese; or (c)
$Normal(\ensuremath{-0.05},
0.02)$, which assumes
an unusually precise prior information about the effect. 
With such unreasonable priors, 
we will get unreasonable posteriors; see the
bottom row of Figure \ref{fig:diffpriors}. This is because
as we increase the precision of the priors we have more
influence on the posterior distribution. Figure
\ref{fig:diffpriors} and Table~\ref{tab:table-priors} illustrate 
how the prior influences the posterior.

At this point, the reader may well ask: why do I have to decide
on a ``reasonable'' prior? What is reasonable anyway? Isn't this injecting 
an uncomfortable level of subjectivity into the analysis? Here, one should 
consider that the way we actually reason about research follows this 
methodology, albeit informally. When we review the literature on a particular
topic,
we report some pattern of results, often classifying them as ``significant''
and ``non-significant'' effects. For example, if we are
reviewing the literature on English relative clauses, we might conclude that
most studies have shown a subject relative advantage. However, if we stop to
consider what the average magnitude of the reported effects is, we already 
have much more information than the binary classification of significant or
not significant. For  the relative clause example, in self-paced reading studies, 
at the critical region (which is the relative clause verb in English), 
we see 67 milliseconds  (SE approximately 20) \citep{grodner}; 
450 ms, 250 ms, 500 ms, and 200 ms (approximate SE 50 ms) in experiments 1-4 
respectively of \cite{gordon01}; 20 ms in \cite{king1991individual} 
(their figure 6). In eye-tracking studies reporting first-pass reading 
time during reading, 
we see 48 ms (no information provided to derive standard error) 
in \cite{staub2010eye}; and
12 ms (no SE provided) in \cite{traxler2002processing}.
Normally we pay no attention to this information when conducting a
new analysis; but using this information is precisely what the Bayesian framework 
allows us to  do. In effect, it allows us to formally build on what we already know.

\begin{table}[h!]
\centering
\caption{Summary of posterior distributions of the coefficients
for the object relative advantage in the Gibson and Wu data, 
assuming different priors. The first three priors can be considered
weakly informative and reasonable, but the last three are overly constrained
and we can see that as a consequence they dominate the posterior,
in the sense that the posterior is largely determined by the 
prior.}
\label{tab:table-priors} 
\begin{tabular}{l r r r r}
\hline
Prior & \multicolumn{2}{c}{95\% CrI} &  $P(\hat{\beta})<0$ & $\hat{\beta}$ \\
\hline
$ Normal(0, 1) $ & $ -0.1 $ & $ 0.02 $ & $ 0.88 $ & $ -0.04 $ \\ $ Normal(0, 0.21) $ & $ -0.09 $ & $ 0.02 $ & $ 0.88 $ & $ -0.03 $ \\ $ Normal(0, 0.11) $ & $ -0.08 $ & $ 0.02 $ & $ 0.86 $ & $ -0.03 $ \\ $ Normal(-0.18, 0.02) $ & $ -0.2 $ & $ -0.15 $ & $ 1 $ & $ -0.17 $ \\ $ Normal(0.05, 0.02) $ & $ 0.01 $ & $ 0.06 $ & $ 0 $ & $ 0.04 $ \\ $ Normal(-0.05, 0.02) $ & $ -0.07 $ & $ -0.02 $ & $ 1 $ & $ -0.04 $ \\
\hline
\end{tabular}
\end{table}

\subsection{Inference using the credible interval}
So we have checked that the posterior is not too sensitive to different
weakly informative priors. What inferences can we draw from the model? 
Are object relative clauses easier than subject relative
clauses in Chinese?   We do have some evidence,
although it is rather weak. Although we do not need to make an
accept/reject decision, for situations where we really want to 
make a decision, \citet{KruschkeEtAl2012} suggest that, since the 95\% credible intervals created
using HDPI include the most credible values of the parameter, they can be used
as a decision tool: One simple decision rule is that any value outside the
95\% HDPI is rejected (also see
\citet{dienes2011bayesian}). Note also that in symmetric posterior 
distributions, the percentile interval will have a range similar to the HPDI
and can equally well be used. 

A more sophisticated decision rule according to
\citet{KruschkeEtAl2012} also allows us to accept  a null result. A region
of practical equivalence (ROPE) around the null value can be established; we
assume that values in that interval, for example $[-0.005,0.005]$, are
practically zero. We would reject the null value if the 95\% HPDI falls
completely outside the ROPE (because none of the most credible values is
practically equivalent to the null value). In addition, we would  accept the
null value if the 95\% HPDI is completely inside the ROPE, because  the  most
credible values are practically equivalent to the null value. The crucial
thing is that 95\% HPDI gets narrower as the sample size gets
larger.

\subsection{Reporting the results of the Gibson and Wu analysis}

So how can we report the analysis of the \citeauthor{gibsonwu} experiment,
and what can we conclude from it? After providing
all the relevant details about the linear mixed model, including the
priors, 
we would report the mean of the estimate of the effect, its credible
interval, and the probability of a negative effect.

In the present case, we would report that (i) the prior for the intercept is 
$Normal(\mu=0,\sigma=10)$, (ii) the prior for the effect of interest (the object-subject difference) is $Normal(0,1)$, and (iii) the regularization on the covariance matrix of random 
effects is $2$. We would also report that 
four chains were run for 2000 iterations each. We would also mention that
a sensitivity analysis using weakly informative priors showed that the 
posterior is not overly influenced by the prior specification. 

The 
results of the \texttt{stan\_lmer} based analysis are 
repeated below for convenience (see 
Table~\ref{tab:lmerstancomparison} for a comparison with \texttt{lmer}
output). If the variance components and correlations of the random effects 
are also of theoretical interest (this could be the case if individual 
differences are relevant theoretically), then the credible intervals
for these can also be reported.

\begin{table}[!htbp]
\caption{Summary of the Bayesian linear mixed model estimates for the 
Gibson and Wu data-set. The stan\_lmer packages provides
the median and the standard deviation of the median absolute 
difference (MAD) of the fixed effects, 
but one could equally well compute the mean and 
standard error, to mirror the lme4 convention.}
\begin{center}
\begin{tabular}{lccc}
\multicolumn{4}{c}{\underline{Random effects}}\\
 \hline
 Groups &  Name     &  Std.Dev. & Corr\\    
 \hline
 subj   &  (Intercept)& 0.2425      & \\
        &  cond       &   0.0762 & -0.521\\ 
 item   &  (Intercept)&     0.1829 &  \\     
        &  cond       &     0.0475 & 0.012\\ 
Residual &            &     0.5131 &  \\
\hline
\multicolumn{4}{c}{\underline{Fixed effects}}\\
\hline
           & Median &  MAD-SD &\\
\hline           
(Intercept) &  6.0641 & 0.0658 &\\
cond        &-0.0364  & 0.0301 &\\
\hline
\end{tabular}
\end{center}
\label{tab:stanresult}
\end{table}%

The effect of interest is the difference between the object and subject
relative clause reading times, and this can be summarized in terms of the
estimated mean of the posterior, and the credible intervals:
($\hat\beta=\ensuremath{-0.04}$, $95\%$
$CrI=[\ensuremath{-0.1},0.02]$). The posterior 
probability of this effect being less than zero, $P(\hat\beta)<0$,
is $0.89$. It has the correct sign following Gibson and Wu's
prediction, but the evidence that it is negative is not very strong.

It would also be very helpful to the reader of a published result to 
have access to the original data and the code that led to the analysis; this
allows the researcher to independently check the analyses himself or herself,
possibly with different prior specifications,
and to build on the published work by using the information gained from the 
published result. Trying out different priors would be specially helpful 
to the expert researcher who has a different opinion (based on their own
knowledge about the topic) on what the true effect might be.

\subsection{Hypothesis testing using the Bayes factor}

The Bayes factor (BF) provides a way to quantify the evidence for the model under which
the observed data are most likely relative to another model. This is accomplished
by computing the ratio of the \emph{marginal likelihoods} of two models $M_0$ and $M_1$, which correspond to research hypotheses
$H_0$ and $H_1$ (we will use the words model and hypothesis 
interchangeably below): 

\begin{equation}
BF_{01} = \frac{p(D|M_0)}{p(D|M_1)}
\end{equation}

$BF_{01}$ then indicates the extent to which the data supports $M_0$ over $M_1$.
The marginal likelihood of a model, $p(D|M)$, is the probability of the
data $D$ given the model $M$. For example, if we toss a coin five times and 
get four heads, we can compute the probability of getting four
heads by using the probability mass function for the binomial distribution:

\begin{equation}
{n \choose k} p^k (1-p)^{(n-k)}
\end{equation}

Here, we have five trials (n=5), four heads (k=4), and 
some probability  $p$ of gettings a heads. 
If the parameter $p$ of the binomial
is believed to be $0.5$, we can compute the
the probability of getting exactly four heads: 

\begin{equation}
{5 \choose 4} 0.5^4 (1-0.5)^{(5-4)} 
= 0.16
\end{equation}

\noindent
This is the marginal likelihood under
a particular model (the assumption that the parameter $p=0.5$). 

As mentioned above, 
the Bayes factor is a ratio: if we want to compare 
the null
hypothesis, $H_0$, with a specific alternative
hypothesis, $H_1$, the ratio we would compute is
$BF_{01} = p(D|H_0)/p(D|H_1)$.
An outcome smaller than one will
mean more evidence for $H_1$ than for $H_0$. Crucially, the Bayes factor can
provide evidence  in favor of the null hypothesis: This is  the case when
the outcome of the calculation is bigger than one (see 
\citet{Gallistel2009}). For example, suppose our null hypothesis is that
our coin is fair, i.e., $p=0.5$, and the alternative is that $p=0.8$. For the 
five coin tosses with four successes, the marginal likelihoods under the two 
hypotheses are  0.16 (for p=0.5), and 
0.41 (for p=0.8). If we take the ratio
of these two values, then we have a Bayes factor of 0.38.
This is weak evidence in 
favor of the alternative hypothesis that $p=0.8$. 
Alternatively, if we had two heads in five tosses, then
the situation would have been different: the marginal likelihoods under the two 
hypotheses would then be  0.31 (for p=0.5), 
and  0.05 (for p=0.8), and the Bayes factor 
would be 6.1, whick is weak evidence in favor of the 
null hypothesis. A scale has been proposed to interpret Bayes factors according 
to the strength of evidence in favor of one model (corresponding to 
some hypothesis) over another (see
\citet{LeeWagenmarks2014}, citing
\citet{jeffreys1961theory}). On this scale, 
a Bayes factor of 10-30 would constitute strong evidence in favor of Model 1
over Model 2; larger values than 10 are very strong evidence, and smaller values
constitute weaker evidence. Obviously, values smaller than 1 would then favor
Model 2.

These examples with coin tosses are simple, but the approach can be scaled
up to the situation where we have a prior defined for our parameter(s);
in this case, the marginal likelihood would be computed by taking a 
sum over the likelihoods, weighted by the probability assigned to each possible value of the parameter.
To take a simplified example, assume that our prior is in the five-trial 
coin-toss example above  is that $p=0.1$ with probability $0.4$ and 
$p=0.8$ with probability $0.6$, then the marginal likelihood when we have
four heads is:

\begin{equation}
0.4\times {5 \choose 4} 0.1^4 (1-0.1)^{(5-4)}+
0.6\times {5 \choose 4} 0.8^4 (1-0.8)^{(5-4)}
\end{equation}

As discussed earlier, in reasonably large samples, the posterior distribution 
is not 
overly influenced by weakly informative priors. In constrast, the Bayes
factor \emph{is} sensitive to the priors \citep{LiuAitkin2008}. When priors are 
defined to allow a broad range of values, the result will be a lower marginal 
likelihood (which in turns influences
the Bayes factor, as we saw in the examples above).
This sensitivity of the Bayes factor to priors can be considered a liability 
\citep[chapter 7.5]{LeeWagenmarks2014}. However, the
dependency on the prior can be studied explicitly with a sensitivity analysis,
in which one varies the prior and studies the fluctuations of the Bayes
factor.

A challenge with the Bayes factor is that, when 
sample sizes are moderate or the models are relatively
complicated, the 
marginal likelihood is often quite difficult to estimate
using sampling  \citep[196]{carlin2008bayesian}. 
Nevertheless, there are tools for computing
the Bayes factor corresponding to 
t-tests \citep{RouderEtAl2009}, and 
the R package \texttt{BayesFactor} and the JASP software 
package also provide functions for computing Bayes factor
for repeated measures ANOVA designs.
There is also another method, called the Savage-Dickey density ratio method that can be used directly with Bayesian linear
mixed models.
We present below a practical example of computing Bayes factor using this method. 

\subsection{An example: Computing Bayes factor in the Gibson and Wu data}

The Savage--Dickey density ratio
method \citep{DickeyLientz1970} is a straightforward way  to compute
the Bayes factor for nested models. The method consists of dividing the
height of the posterior for a certain parameter by the height of the prior of the
same parameter, at the point of interest (see
\citet{WagenmakersEtAl2010} for a complete tutorial and the mathematical
proof). Critically, we can use the height of an approximation of the posterior
distribution from the samples obtained from the numerical method employed
(such as MCMC). In our case, we could calculate the evidence in favor or
against our predictor (condition) being zero
The model with the experimental condition and the null model will have several
parameters in common that are not of interest (such as the intercept of the
fixed effects and random effects, the standard deviation, etc.), but these
parameters won't influence the calculation of the Savage--Dickey density ratio
\citep{WagenmakersEtAl2010}.

\singlespacing
\begin{listing}
\begin{knitrout}
\definecolor{shadecolor}{rgb}{1, 1, 1}\color{fgcolor}\begin{kframe}
\begin{Verbatim}[numbers=left,fontfamily=courier,fontsize=\footnotesize, firstnumber=last]
library(polspline)
 \end{Verbatim}
\begin{Verbatim}[numbers=left,fontfamily=courier,fontsize=\footnotesize, firstnumber=last]
fit_posterior <- logspline(posterior_condition)
 \end{Verbatim}
\begin{Verbatim}[numbers=left,fontfamily=courier,fontsize=\footnotesize, firstnumber=last]
posterior <- dlogspline(0, fit_posterior) # Height of the posterior at 0
 \end{Verbatim}
\begin{Verbatim}[numbers=left,fontfamily=courier,fontsize=\footnotesize, firstnumber=last]
prior     <- dnorm(0, 0, 1) # Height of the prior at 0
 \end{Verbatim}
\begin{Verbatim}[numbers=left,fontfamily=courier,fontsize=\footnotesize, firstnumber=last]
(BF01 <- posterior/prior) #BF01 shows clear support for H0 
 \end{Verbatim}
\begin{Verbatim}[fontfamily=courier,fontsize=\footnotesize, firstnumber=last]
## [1] 15.67
 \end{Verbatim}
\end{kframe}
\end{knitrout}
\caption{Code for calculating the Bayes factor for the Gibson and Wu 
data.}\label{code:BF}
\end{listing}

Listing \ref{code:BF} illustrates how to perform the calculation for
the Gibson and Wu example. We see
that the comparison clearly favors the null hypothesis: it is showing
15.67 times more evidence for the null than for any other value.
However, this might be because when priors allow a broad range of values, 
and thus are too
uninformative, the alternative hypothesis to the null, $H_1$, (that the effect
is different from zero) is penalized for assigning too much prior mass to
values that are too unlikely (while all the prior mass of the null hypothesis,
$H_0$, is concentrated in zero). Without a proper specification of priors,
$H_0$ would always be  more likely than $H_1$.

Table \ref{tab:BFs} shows the Bayes factor under different weakly informative
priors: the first column represents the numerator of a Bayes factor, while the
second column the denominator. The priors in Table \ref{tab:BFs} represent our
prior beliefs on the plausibility of different values of the effect of object
vs.\ subject relative clauses. The table shows that as we provide tighter and
more realistic priors, the evidence in favor of $H_0$ decreases, which means
that we don't have enough evidence to accept the null hypothesis. 
But notice that this doesn't mean that we can accept $H_1$ either.

\begin{table}[h!]
\centering
\caption{Bayes factor under different weakly informative priors.}
\begin{tabular}{r r r }
\hline
H0 & H1 & Prior \\
\hline
3.63 & 1 & Normal(0, 0.21) \\ 2.14 & 1 & Normal(0, 0.11) \\
\hline
\end{tabular}
\label{tab:BFs} 
\end{table}

In sum, the Bayes factor can be a useful tool, but it should be borne in mind
that it will always be affected by the prior, so a sensitivity analysis
is a good idea when reporting the Bayes factor. 
According to
\citet{dienes2011bayesian}, its calculation  depends on answering a question
about which there may be disagreement among researchers:
``What way of assigning probability distributions of effect sizes as predicted
by theories would be accepted by protagonists on all sides of a debate?''
One of the clearest advantages of the Bayes factor is that once the minimal
magnitude of an expected effect is agreed upon, evidence can be gathered in favor
of the null hypothesis. 

\subsection{Model selection using cross-validation}

Another way to make a decision about the hypothesis that object relative
clauses are easier than subject relative clauses in Chinese is to treat the
hypothesis as a model that can be compared with other models such as the null.
We will focus on 
cross-validation; for other approaches, see \cite{ShiffrinEtAl2008}.


The question whether  object relative clauses are easier than subject relative
clauses in Chinese can be also be phrased in terms of evaluating the model on
its ability to make predictions about future or unseen observations, in
comparison with, for example, a null model (or another model). 
However, it may not be the most suitable
way to compare nested linear mixed models when the effects being investigated
are small \citep{WangGelman2014difficulty,GelmanEtAl2014understanding}.

This approach to model selection is based on finding the most ``useful model''
for characterizing future data, and  not necessarily the true model:  the true
model is not guaranteed to produce the best predictions, and a false model is
not guaranteed to produce poor predictions \citep{WangGelman2014difficulty}.
The ideal measure of a model's fit would be its (out-of-sample) predictive
performance for new observations that are produced by the same data-generating
process. When the future observations are not available the
predictive performance can be estimated by calculating the \emph{expected
predictive performance} 
\citep{GelmanEtAl2014understanding,VehtariOjanen2012}. 

The cross-validation
techniques that we review below are based on comparing the expected predictive
performance of a model with its actual performance (but see
\citet{VehtariOjanen2012} and \citet{PiironenVehtari2015} for a more
complete review). We will focus on Bayesian leave-one-out cross-validation (LOO-CV;
\citet{GeisserEddy1979}) and three approximations: 
(a) k-fold-cross-validation
\citep[k-fold-CV;][]{VehtariOjanen2012}, 
(b) Pareto smoothed importance sampling
\citep[PSIS-LOO;][]{VehtariGelman2015Pareto}, and  
(c) the widely applicable
information criterion (or Watanabe-Akaike information criterion: WAIC;
\citet{Watanabe2009,Watanabe2010}). The latter two 
are implemented in the R package \emph{loo} \citep{loo}, but they should be
used with care, since they are affected by highly influential observations.
When highly influential observations are present, k-fold-CV is recommended
\citep[the code for implementing k-fold-CV in Stan is available
in][]{VehtariEtAl2015Loo}.

The basic idea of cross-validation is to split the data such that
each subset is used as a validation set, while the the remaining sets (the
training set) are used for estimating the parameters. LOO-CV method depicts the
case when the training set only excludes one observation. The main advantage
of this methods is its robustness, since the training set is as similar as
possible to the real data, while  the same observations are never used
simultaneously for training and evaluating the predictions. A major
disadvantage is the computational burden \citep{VehtariOjanen2012}, since
we need to fit a model as many times as the number of observations.

The k-fold-CV \citep{VehtariOjanen2012} can be used to reduce the computation
time by reducing the number of models we need to fit. In the k-fold-CV
approach, the data are split into k subsets (or folds), where k is generally
around ten. Each subset is in turn used as the validation set, while the
remaining data are used for parameter estimation. A further reduction in
computation time can be achieved with PSIS-LOO
\citep{VehtariGelman2015Pareto}, which is faster compared to
LOO-CV, and does not require fitting the model multiple times 
\citep{VehtariGelman2015Pareto,VehtariEtAl2015Loo}.


Information criteria are commonly used for selecting Bayesian models, since
they are directly related to assessing the predictive performance of the
models. In addition, WAIC is asymptotically equal to LOO \citep{Watanabe2010}.
The distinguishing feature of WAIC in comparison with AIC (Akaike Information
Criterion; \citet{Akaike1974}), DIC (Deviance Information Criterion;
\citet{SpiegelhalterEtAl2002}), and BIC (Bayesian Information Criterion;
\citet{Schwarz1978}, which also has a different goal than the other
measures discussed here), is that WAIC is point-wise: the uncertainty is
calculated point-by-point in the data over the entire posterior distribution
\citep{GelmanEtAl2014understanding}. This is important because some
observations are harder to predict than others. In addition, AIC does not work
well with strong priors, and while DIC can take into account informative
priors, and it is the measure of choice in many Bayesian applications, 
it may
give unexpected results when the posterior distribution is not well
summarized by its mean (for example, if the posterior is substantially
skewed). For a complete comparison between AIC, DIC, and WAIC, see
\citep{GelmanEtAl2014understanding}. 

Cross validation techniques are ideally suited for comparing highly different
models, and may be a fully Bayesian replacement for AIC or DIC. However, even
with  moderate sample size, it can be difficult to compare nested
hierarchical models (such as linear mixed models) based on predictive accuracy
\citep{WangGelman2014difficulty}. An experimental manipulation can produce a
tiny change in predictive accuracy, which can be nearly indistinguishable from
noise, but it can be still useful for evaluating a psycholinguistic theory.
This means that unless we are dealing with huge effects or with a very large
sample size,  the null model would  always be almost as good as the model with
the predictor of interest, as far as predictive accuracy is concerned, while
the complexity of the model with the predictor of interest is penalized
\citep{WangGelman2014difficulty,GelmanEtAl2014understanding}.

\subsubsection{Some closing remarks on inference}

In the previous sections, we have presented Bayesian methods as a 
useful and important
alternative to null hypothesis significance testing (NHST) 
for statistical inference. In contrast to NHST, where a
sharp binary decision is made between 
rejecting or failing to reject the null hypothesis,
we can now directly talk about the strength of the evidence for a certain effect. We pointed out that the 95\% Bayesian credible interval as a possible way to summarize
the evidence. The credible interval  has  a very intuitive interpretation
(which researchers often ascribe mistakenly to frequentist confidence
intervals): it gives the range over which we can be 95\% certain that the true
value of the effect lies, given of course the data, the priors,
and the model. We 
also pointed out that the mass of the probability of the posterior distribution below
(or above) zero can give valuable information about the plausibility of a
negative (or positive) effect.

As a rule of thumb, we can interpret the evidence as strong if zero lies
outside the 95\% credible interval \citep{KruschkeEtAl2012}. If zero is
included within the interval, there might still be weak evidence for an
effect, if the probability of the estimate being less than (or greater than)
zero is large enough. Our interpretation of the evidence should also take into
account that the range of possible magnitudes of the effect makes sense
theoretically; for example,  if we find an effect of less than one millisecond, we
shouldn't interpret it as strong evidence, just because its credible interval
doesn't include zero. A difference of less than one millisecond 
would likely have no theoretical relevance in psycholinguistics or linguistics.

Given the results from the Bayesian analysis of the Gibson and Wu data, 
we could claim that 
there is some weak evidence for the claim that object relative clauses are easier
than subject relative clauses in Chinese, depending on what we make of the
magnitude of the effect on this experiment in comparison with other similar
experiments. 
Importantly, we wouldn't
be able to claim that we have evidence for no effect. This is a common
problem in the way that null hypothesis significance testing 
is used in linguistics and psycholinguistics; 
a failure to find an effect is presented as evidence
for the null hypothesis that the parameter is 0 (also see Part I of this
review for more discussion).
The effect is also considered to be zero even if repeated experiments
consistently 
show, say, a negative sign of the effect that do not reach statistical 
significance.
In such a situation, if theory suggests ``no effect'', we
could  (and should) establish an interval around zero that would be
practically equivalent to ``no effect'', the region of practical equivalence
or ROPE, and find that the 95\%
credible interval of the estimate of the effect  falls completely inside it.
Alternatively, one could identify the smallest effect we would expect, 
and then use Bayes factors.

%

In the final section below, we review some examples that use 
Bayesian tools, and mention some of the possibilities for fitting more 
complex and interesting models. Any of these example applications can 
serve as a starting point for the researcher.

\section{Examples of applications of Bayesian Methods}

It has become relatively straightforward to fit complex Bayesian models 
due to the increase in computing power and 
the appearance of probabilistic programming languages, such as WinBUGS
\citep{lunn2000winbugs}, JAGS \citep{plummer2011jags}, and Stan
\citep{Stan2015}. Even
though these statistical packages allow the user to define models without
having to deal with the complexities of the sampling process, some background
statistical knowledge is needed before one can define the models.

There are some alternatives that allow Bayesian inference in R without having
to fully specify the model ``by hand''. The packages \emph{rstanarm}
\citep{rstanarm2016} and \emph{brms} \citep{brms2016} emulates many popular R
model-fitting functions, such as (g)lmer, using  Stan for the back-end
estimation and sampling, and can be useful for a smooth transition between frequentist
linear mixed models and Bayesian ones. In addition, the \emph{BayesFactor}
\citep{morey2015bayesfactor} package emulates other standard frequentist tests
(t-test, ANOVA, linear models, etc.), and provides the Bayes factor given some
pre-specified priors. For a simpler option, JASP \citep{JASP2015}
provides a graphical user interface, and is an alternative to SPSS.

For linear mixed models,  one  strength of Bayesian methods is that we can fit
models with a full random structure that would not converge with frequentist
methods or would yield overestimates of correlations between the random
effects \citep{BatesEtAlParsimonious}. This can be achieved by using appropriate weakly informative
priors for the correlation matrices (so-called LKJ priors, and see also
\citet{SorensenVasishthTutorial} for a tutorial).
Some examples of papers using Bayesian
linear mixed models in psycholinguistics are
\citet{FrankEtAl2015,HofmeisterVasishth2014,HusainEtAl2014}. However, the
major advantage of Bayesian methods lies in the possibility of moving beyond
linear models. These become relevant for modeling distributions of reaction and reading times (RTs), which are limited
on the left by some amount of time (i.e., the shift of the distribution), and
are highly right skewed. RTs can be reciprocal- or log-transformed  to
incorporate them in linear mixed models, but these transformations still
assume a shift of 0 ms. \citet{Rouder2005} suggests then the shifted
log-normal hierarchical model as a suitable model for RTs. This type of model
is not linear but can be fit straightforwardly,
and can be used for inferences in experiments with self-paced reading tasks
\citep{NicenboimEtAlFrontiers2015Capacity}. Another potential use of
non-linear hierarchical models that to our knowledge has not been applied in
psycholinguistics or linguistics is ordered probit hierarchical models for
acceptability judgments or any type of rating task that uses a scale
\citep[Chapter 23]{kruschke2010doing}. 
A further interesting application is that one can synthesize evidence from
existing studies by carrying out a meta-analysis 
\citep{VasishthetalPLoSOne2013}. Meta-analysis is not widely used in 
linguistics and psycholinguistics, but it can serve a very important
role in literature reviews.

Bayesian cognitive modeling is another extremely fruitful use of Bayesian
methods, and some of the methods discussed in \citet{Lee2011} and
\citet{LeeWagenmarks2014} could easily be adapted for psycholinguistics. It is
important to note the distinction between using Bayesian methods for
modeling cognitive processes, assuming that (some aspect of) the 
mind is Bayesian, and using Bayesian methods for modeling 
cognitive processes without necessarily assuming a Bayesian mind.
Some examples of the former category are Bayesian/noisy channels approaches to parsing (for a review see \citet{Traxler2014}) or to word learning (see, for example, \citet{XuTenenbaum2007}); and the belief update models presented by \citet{MyslinLevy2016} and \citet{KleinschmidtEtAl2012}. Indeed, even though \citet{KleinschmidtEtAl2012}
model adaptation as a Bayesian belief update, the model itself was fit
using frequentist methods.
An example of the second category, i.e.,  using Bayesian methods for modeling without
assuming that the mind is Bayesian, is \citet{LogacevVasishthQJEP2015InPress}.
Here, the focus is on evaluating
different models of parsing in the face of ambiguities.

In addition, there is a class of models that is mostly used  in
two-forced choice tasks and that have the strength of integrating accuracy and 
reaction times instead of wrongly treating them as independent outcomes. This
class of models is based on the idea that response selection can be modeled by 
a process that accumulates evidence until a threshold in reached. This could be 
applied in deciding whether a string of letters is a word or a non-word, whether 
a word is the right completion of a sentence, whether a sentence is
grammatical or not, and so forth. One of the most widely  applied evidence
accumulation model is the Ratcliff diffusion model (see
\citet{RatcliffRouder1998}), but  several  other models based on the similar
ideas exist, such as the Ballistic and linear Ballistic accumulator
\citep{BrownHeathcote2005,BrownHeathcote2008}.
The implementation of these models with frequentist methods is
notoriously complicated, and in order to fit all their parameters they require
a large number of trials. Bayesian methods allow extending these models
hierarchically, with all the benefits that this implies, that is, the ability
to take into account within- and 
between-subjects and between-items variability, and to do
partial pooling. \citet{VandekerckhoveTuerlinckx2011}, for example, provide an
implementation in WinBUGS of a hierarchical diffusion model. The linear
ballistic accumulator has also been implemented in WinBUGS
\citep{DonkinEtAl2009}, and it could be extended in the same way as the
hierarchical diffusion model was. Another Bayesian model based on the
accumulation of evidence is the lognormal race model \citep{RouderEtAl2014},
not as feature rich as the diffusion model and the linear ballistic
accumulator, but its approach can generalize to any number of choices
(including just one choice).

Finally, another advantage that Bayesian methods can provide is the use of
informative priors in situations where data are scarce but we have previous
knowledge about the effects. This is the idea behind  the Small N Acceptability
Paradigm for Linguistic Acceptability Judgments (SNAP Judgments;
\citet{MahowaldEtAlUnpublished}), which allows us to obtain quantitative and
statistically valid data in syntax and semantics research in situations where
it would be difficult to consult with many native speakers. And finally,  the
use of informative priors could be a significant advantage in studies
with impaired participants (such as aphasics), where it is difficult to have a 
large sample size.

\section{Concluding remarks}

Carrying out Bayesian data analysis clearly requires thought and effort; 
even if one uses convenient packages like \texttt{rstanarm},
several decisions have to be made: we have to define 
priors, carry out sensitivity analyses, and decide how to interpret the results.
By comparison, fitting a linear mixed model using \texttt{lme4} is much easier:
just write a single line of code and extract the t-value(s) or the like.
To add insult to injury, the overhead in terms of time and effort of fitting a 
Bayesian model seems unjustified given that, 
for large sample sizes, the estimates for the fixed effects
from a Bayesian model and the corresponding \texttt{lme4} model will be quite 
similar (if not identical), especially  with weakly informative priors 
(for examples from psycholinguistics, see \citet{BatesEtAlParsimonious}).
Why bother to use Bayesian methods then? One compelling reason is that although
p-values answer \textit{a} question, they answer the wrong question. 
Once one realizes that the p-value doesn't provide any direct evidence for 
the research question, the motivation to compute it fades. Another reason is that
since 
we already tend to interpret the result of frequentist analyses in a Bayesian 
manner, we might as well carry out a Bayesian analysis. Finally, 
as discussed earlier, Bayesian probabilistic programming languages provide a degree of flexibility in defining models that is difficult to match with frequentist tools.

\section{Further reading}

For a first introduction to Bayesian methods we suggest
\citet{mcelreath2015statistical} and \citet{kruschke2010doing}. 
\citet{lynch2007introduction} is also excellent but assumes some calculus.
For a more advanced treatment of the topic,
see \citet{Gelman14}. 
Linear mixed models are covered from both the frequentist and Bayesian perspective by \citet{gelmanhill07}.
For an accessible introduction of Bayesian methods for cognitive modeling see \citet{LeeWagenmarks2014}.

\section*{Acknowledgment}
Thanks to Lena J\"ager, Dario Paape, and Daniela Mertzen for helpful comments on previous versions of this review. 

\bibliography{LLC}

\begin{thebibliography}{}

\bibitem[Akaike, 1974]{Akaike1974}
Akaike, H. (1974).
\newblock A new look at the statistical model identification.
\newblock {\em {IEEE Transactions on Automatic Control}}, 19(6):716--723.

\bibitem[Barr et~al., 2013]{barr2013random}
Barr, D.~J., Levy, R., Scheepers, C., and Tily, H.~J. (2013).
\newblock Random effects structure for confirmatory hypothesis testing: Keep it
  maximal.
\newblock {\em Journal of Memory and Language}, 68(3):255--278.

\bibitem[Bates et~al., 2015a]{BatesEtAlParsimonious}
Bates, D., Kliegl, R., Vasishth, S., and Baayen, H. (2015a).
\newblock Parsimonious mixed models.
\newblock ArXiv e-print.

\bibitem[Bates et~al., 2015b]{lme4new}
Bates, D., Maechler, M., Bolker, B., and Walker, S. (2015b).
\newblock Fitting linear mixed-effects models using lme4.
\newblock {\em Journal of Statistical Software}.
\newblock In Press.

\bibitem[Bates and Sarkar, 2007]{batessarkar}
Bates, D. and Sarkar, D. (2007).
\newblock {\em lme4: {L}inear mixed-effects models using {S4} classes}.
\newblock R package version 0.9975-11.

\bibitem[Box and Tiao, 1992]{box2011bayesian}
Box, G.~E. and Tiao, G.~C. (1992).
\newblock {\em Bayesian inference in statistical analysis}.
\newblock John Wiley \& Sons, first edition.

\bibitem[Brown and Heathcote, 2005]{BrownHeathcote2005}
Brown, S.~D. and Heathcote, A. (2005).
\newblock A ballistic model of choice response time.
\newblock {\em Psychological review}, 112(1):117.

\bibitem[Brown and Heathcote, 2008]{BrownHeathcote2008}
Brown, S.~D. and Heathcote, A. (2008).
\newblock The simplest complete model of choice response time: {L}inear
  ballistic accumulation.
\newblock {\em Cognitive psychology}, 57(3):153--178.

\bibitem[Buerkner, 2015]{brms2016}
Buerkner, P.-C. (2015).
\newblock {\em brms: Bayesian Regression Models using {S}tan}.
\newblock R package version 0.6.0.

\bibitem[Carlin and Louis, 2008]{carlin2008bayesian}
Carlin, B.~P. and Louis, T.~A. (2008).
\newblock {\em Bayesian methods for data analysis}.
\newblock CRC Press.

\bibitem[Chung et~al., 2013]{chung2013weakly}
Chung, Y., Gelman, A., Rabe-Hesketh, S., Liu, J., and Dorie, V. (2013).
\newblock Weakly informative prior for point estimation of covariance matrices
  in hierarchical models.
\newblock {\em Manuscript submitted for publication}.

\bibitem[Dickey et~al., 1970]{DickeyLientz1970}
Dickey, J.~M., Lientz, B., et~al. (1970).
\newblock The weighted likelihood ratio, sharp hypotheses about chances, the
  order of a markov chain.
\newblock {\em The Annals of Mathematical Statistics}, 41(1):214--226.

\bibitem[Dienes, 2011]{dienes2011bayesian}
Dienes, Z. (2011).
\newblock Bayesian versus orthodox statistics: Which side are you on?
\newblock {\em Perspectives on Psychological Science}, 6(3):274--290.

\bibitem[Donkin et~al., 2009]{DonkinEtAl2009}
Donkin, C., Averell, L., Brown, S., and Heathcote, A. (2009).
\newblock Getting more from accuracy and response time data: {M}ethods for
  fitting the linear ballistic accumulator.
\newblock {\em Behavior Research Methods}, 41(4):1095--1110.

\bibitem[Frank et~al., 2015]{FrankEtAl2015}
Frank, S.~L., Trompenaars, T., and Vasishth, S. (2015).
\newblock Cross-linguistic differences in processing double-embedded relative
  clauses: {W}orking-memory constraints or language statistics?
\newblock {\em Cognitive Science}, page n/a.

\bibitem[Gabry and Goodrich, 2016]{rstanarm2016}
Gabry, J. and Goodrich, B. (2016).
\newblock {\em rstanarm: Bayesian Applied Regression Modeling via Stan}.
\newblock R package version 2.9.0-1.

\bibitem[Gallistel, 2009]{Gallistel2009}
Gallistel, C.~R. (2009).
\newblock The importance of proving the null.
\newblock {\em Psychological Review}, 116(2):439–453.

\bibitem[Geisser and Eddy, 1979]{GeisserEddy1979}
Geisser, S. and Eddy, W.~F. (1979).
\newblock A predictive approach to model selection.
\newblock {\em Journal of the American Statistical Association},
  74(365):153--160.

\bibitem[Gelman, 2006]{gelman2006prior}
Gelman, A. (2006).
\newblock Prior distributions for variance parameters in hierarchical models
  (comment on article by {Browne and Draper}).
\newblock {\em Bayesian analysis}, 1(3):515--534.

\bibitem[Gelman et~al., 2014a]{Gelman14}
Gelman, A., Carlin, J.~B., Stern, H.~S., Dunson, D.~B., Vehtari, A., and Rubin,
  D.~B. (2014a).
\newblock {\em Bayesian Data Analysis}.
\newblock Chapman and Hall/CRC, third edition.

\bibitem[Gelman and Hill, 2007]{gelmanhill07}
Gelman, A. and Hill, J. (2007).
\newblock {\em Data analysis using regression and multilevel/hierarchical
  models}.
\newblock Cambridge University Press, Cambridge, UK.

\bibitem[Gelman et~al., 2012]{GelmanEtAl2012}
Gelman, A., Hill, J., and Yajima, M. (2012).
\newblock Why we (usually) don’t have to worry about multiple comparisons.
\newblock {\em Journal of Research on Educational Effectiveness},
  5(2):189–211.

\bibitem[Gelman et~al., 2014b]{GelmanEtAl2014understanding}
Gelman, A., Hwang, J., and Vehtari, A. (2014b).
\newblock Understanding predictive information criteria for {B}ayesian models.
\newblock {\em Statistics and Computing}, 24(6):997--1016.

\bibitem[Gelman et~al., 2008]{gelman2008weakly}
Gelman, A., Jakulin, A., Pittau, M.~G., and Su, Y.-S. (2008).
\newblock A weakly informative default prior distribution for logistic and
  other regression models.
\newblock {\em The Annals of Applied Statistics}, pages 1360--1383.

\bibitem[Gibson and Wu, 2013]{gibsonwu}
Gibson, E. and Wu, H.-H.~I. (2013).
\newblock Processing {C}hinese relative clauses in context.
\newblock {\em Language and Cognitive Processes}, 28(1-2):125--155.

\bibitem[Gordon et~al., 2001]{gordon01}
Gordon, P.~C., Hendrick, R., and Johnson, M. (2001).
\newblock Memory interference during language processing.
\newblock {\em Journal of Experimental Psychology: Learning, Memory and
  Cognition}, 27(6):1411--1423.

\bibitem[Grodner and Gibson, 2005]{grodner}
Grodner, D. and Gibson, E. (2005).
\newblock Consequences of the serial nature of linguistic input.
\newblock {\em Cognitive Science}, 29:261--290.

\bibitem[Haller and Krauss, 2002]{HallerKrauss2002}
Haller, H. and Krauss, S. (2002).
\newblock Misinterpretations of significance: A problem students share with
  their teachers.
\newblock {\em Methods of Psychological Research}, 7(1):1--20.

\bibitem[Hoekstra et~al., 2014]{Hoekstra2014}
Hoekstra, R., Morey, R.~D., Rouder, J.~N., and Wagenmakers, E.-J. (2014).
\newblock Robust misinterpretation of confidence intervals.
\newblock {\em Psychonomic Bulletin \& Review}, 21(5):1157–1164.

\bibitem[Hofmeister and Vasishth, 2014]{HofmeisterVasishth2014}
Hofmeister, P. and Vasishth, S. (2014).
\newblock Distinctiveness and encoding effects in online sentence
  comprehension.
\newblock {\em Frontiers in Psychology}, 5:1--13.
\newblock Article 1237.

\bibitem[Hsiao and Gibson, 2003]{hsiao03}
Hsiao, F. P.-F. and Gibson, E. (2003).
\newblock Processing relative clauses in {C}hinese.
\newblock {\em Cognition}, 90:3--27.

\bibitem[Husain et~al., 2014]{HusainEtAl2014}
Husain, S., Vasishth, S., and Srinivasan, N. (2014).
\newblock Strong expectations cancel locality effects: {E}vidence from {H}indi.
\newblock {\em PLoS ONE}, 9(7):1--14.

\bibitem[Jeffreys, 1961]{jeffreys1961theory}
Jeffreys, H. (1961).
\newblock {\em Theory of probability}.
\newblock Oxford: Clarendon Press.

\bibitem[King and Just, 1991]{king1991individual}
King, J. and Just, M.~A. (1991).
\newblock Individual differences in syntactic processing: The role of working
  memory.
\newblock {\em Journal of memory and language}, 30(5):580--602.

\bibitem[Kleinschmidt et~al., 2012]{KleinschmidtEtAl2012}
Kleinschmidt, D., Fine, A.~B., and Jaeger, T.~F. (2012).
\newblock A belief-updating model of adaptation and cue combination in
  syntactic comprehension.
\newblock In {\em Proceedings of the {34rd Annual Meeting of the Cognitive
  Science Society (CogSci12)}}, pages 605--10.

\bibitem[Kruschke, 2015]{kruschke2010doing}
Kruschke, J.~K. (2015).
\newblock {\em Doing {B}ayesian Data Analysis}.
\newblock Academic Press, Boston, second edition.

\bibitem[Kruschke et~al., 2012]{KruschkeEtAl2012}
Kruschke, J.~K., Aguinis, H., and Joo, H. (2012).
\newblock The time has come: {Bayes}ian methods for data analysis in the
  organizational sciences.
\newblock {\em Organizational Research Methods}, 15(4):722–752.

\bibitem[Lecoutre et~al., 2003]{LecoutreEtAl2003}
Lecoutre, M.-P., Poitevineau, J., and Lecoutre, B. (2003).
\newblock Even statisticians are not immune to misinterpretations of null
  hypothesis significance tests.
\newblock {\em International Journal of Psychology}, 38(1):37--45.

\bibitem[Lee, 2011]{Lee2011}
Lee, M.~D. (2011).
\newblock How cognitive modeling can benefit from hierarchical {Bayes}ian
  models.
\newblock {\em Journal of Mathematical Psychology}, 55(1):1--7.

\bibitem[Lee and Wagenmakers, 2014]{LeeWagenmarks2014}
Lee, M.~D. and Wagenmakers, E.-J. (2014).
\newblock {\em Bayesian cognitive modeling: A practical course}.
\newblock Cambridge University Press.

\bibitem[Levshina, 2016]{Levshina2016}
Levshina, N. (2016).
\newblock A {B}ayesian mixed-effect multinomial model of {E}nglish permissive
  constructions reveals a remarkable alignment of linguistic, cognitive, social
  and collostructional distances.
\newblock {\em Cognitive Linguistics}.
\newblock In Press.

\bibitem[Liu and Aitkin, 2008]{LiuAitkin2008}
Liu, C.~C. and Aitkin, M. (2008).
\newblock {Bayes} factors: Prior sensitivity and model generalizability.
\newblock {\em Journal of Mathematical Psychology}, 52(6):362–375.

\bibitem[Liu et~al., 2013]{liu2013simulation}
Liu, Y., Gelman, A., and Zheng, T. (2013).
\newblock Simulation-efficient shortest probability intervals.
\newblock {\em arXiv preprint arXiv:1302.2142}.

\bibitem[Loga{\v c}ev and Vasishth, 2016]{LogacevVasishthQJEP2015InPress}
Loga{\v c}ev, P. and Vasishth, S. (2016).
\newblock Understanding underspecification: {A} comparison of two computational
  implementations.
\newblock {\em Quarterly Journal of Experimental Psychology}.
\newblock Accepted.

\bibitem[Love et~al., 2015]{JASP2015}
Love, J., Selker, R., Marsman, M., Jamil, T., Dropmann, D., Verhagen, A.~J.,
  Ly, A., Gronau, Q.~F., Smira, M., Epskamp, S., Matzke, D., Wild, A., Rouder,
  J.~N., Morey, R.~D., and Wagenmakers, E.~J. (2015).
\newblock {JASP} (version 0.7)[computer software].

\bibitem[Lunn et~al., 2000]{lunn2000winbugs}
Lunn, D., Thomas, A., Best, N., and Spiegelhalter, D. (2000).
\newblock {WinBUGS}-{A B}ayesian modelling framework: {C}oncepts, structure,
  and extensibility.
\newblock {\em Statistics and computing}, 10(4):325--337.

\bibitem[Lynch, 2007]{lynch2007introduction}
Lynch, S.~M. (2007).
\newblock {\em Introduction to applied {B}ayesian statistics and estimation for
  social scientists}.
\newblock Springer.

\bibitem[Mahowald et~al., 2015]{MahowaldEtAlUnpublished}
Mahowald, K., Graff, P., Hartman, J., and Gibson, E. (2015).
\newblock Snap judgments: A small n acceptability paradigm (snap) for
  linguistic acceptability judgments.

\bibitem[McElreath, 2015]{mcelreath2015statistical}
McElreath, R. (2015).
\newblock {\em Statistical rethinking: A Bayesian course with R examples}.
\newblock Chapman and Hall/CRC.

\bibitem[Morey et~al., 2015]{MoreyEtAl2015}
Morey, R.~D., Hoekstra, R., Rouder, J.~N., Lee, M.~D., and Wagenmakers, E.-J.
  (2015).
\newblock The fallacy of placing confidence in confidence intervals.
\newblock {\em Psychonomic Bulletin \& Review}.

\bibitem[Morey and Rouder, 2015]{morey2015bayesfactor}
Morey, R.~D. and Rouder, J. (2015).
\newblock Bayesfactor: An r package for bayesian analysis in common research
  designs.

\bibitem[Mysl{\'\i}n and Levy, 2016]{MyslinLevy2016}
Mysl{\'\i}n, M. and Levy, R. (2016).
\newblock Comprehension priming as rational expectation for repetition:
  Evidence from syntactic processing.
\newblock {\em Cognition}, 147:29--56.

\bibitem[Nicenboim et~al., 2015]{NicenboimEtAlFrontiers2015Capacity}
Nicenboim, B., Loga{\v c}ev, P., Gattei, C., and Vasishth, S. (2015).
\newblock When high-capacity readers slow down and low-capacity readers speed
  up: {W}orking memory differences in unbounded dependencies.
\newblock Resubmitted.

\bibitem[O'Hagan et~al., 2006]{ohagan2006uncertain}
O'Hagan, A., Buck, C.~E., Daneshkhah, A., Eiser, J.~R., Garthwaite, P.~H.,
  Jenkinson, D.~J., Oakley, J.~E., and Rakow, T. (2006).
\newblock {\em Uncertain judgements: {E}liciting experts' probabilities}.
\newblock John Wiley \& Sons.

\bibitem[Piironen and Vehtari, 2015]{PiironenVehtari2015}
Piironen, J. and Vehtari, A. (2015).
\newblock Comparison of {B}ayesian predictive methods for model selection.
\newblock {\em arXiv preprint arXiv:1503.08650}.

\bibitem[Plummer, 2012]{plummer2011jags}
Plummer, M. (2012).
\newblock {JAGS} version 3.3.0 manual.
\newblock {\em International Agency for Research on Cancer. Lyon, France}.

\bibitem[Ratcliff, 1993]{Ratcliff1993}
Ratcliff, R. (1993).
\newblock Methods for dealing with reaction time outliers.
\newblock {\em Psychological Bulletin}, 114(3):510.

\bibitem[Ratcliff and Rouder, 1998]{RatcliffRouder1998}
Ratcliff, R. and Rouder, J.~N. (1998).
\newblock {Modeling Response Times for Two-Choice Decisions}.
\newblock {\em Psychological Science}, 9(5):347--356.

\bibitem[Rouder, 2005]{Rouder2005}
Rouder, J.~N. (2005).
\newblock Are unshifted distributional models appropriate for response time?
\newblock {\em Psychometrika}, 70(2):377--381.

\bibitem[Rouder et~al., 2014]{RouderEtAl2014}
Rouder, J.~N., Province, J.~M., Morey, R.~D., Gomez, P., and Heathcote, A.
  (2014).
\newblock The lognormal race: A cognitive-process model of choice and latency
  with desirable psychometric properties.
\newblock {\em Psychometrika}, pages 1--23.

\bibitem[Rouder et~al., 2009]{RouderEtAl2009}
Rouder, J.~N., Speckman, P.~L., Sun, D., Morey, R.~D., and Iverson, G. (2009).
\newblock {Bayes}ian t tests for accepting and rejecting the null hypothesis.
\newblock {\em Psychonomic Bulletin \& Review}, 16(2):225–237.

\bibitem[Schielzeth and Forstmeier, 2009]{SchielzethForstmeier2009}
Schielzeth, H. and Forstmeier, W. (2009).
\newblock {Conclusions beyond support: Overconfident estimates in mixed
  models}.
\newblock {\em Behavioral Ecology}, 20(2):416--420.

\bibitem[Schwarz, 1978]{Schwarz1978}
Schwarz, G. (1978).
\newblock Estimating the dimension of a model.
\newblock {\em The Annals of Statistics}, 6(2):461--464.

\bibitem[Shiffrin et~al., 2008]{ShiffrinEtAl2008}
Shiffrin, R., Lee, M., Kim, W., and Wagenmakers, E.-J. (2008).
\newblock A survey of model evaluation approaches with a tutorial on
  hierarchical {Bayes}ian methods.
\newblock {\em HCOG}, 32(8):1248–1284.

\bibitem[Sorensen et~al., 2015]{SorensenVasishthTutorial}
Sorensen, T., Hohenstein, S., and Vasishth, S. (2015).
\newblock {B}ayesian linear mixed models using {S}tan: {A} tutorial for
  psychologists, linguists, and cognitive scientists.
\newblock ArXiv e-print.

\bibitem[Spiegelhalter et~al., 2002]{SpiegelhalterEtAl2002}
Spiegelhalter, D.~J., Best, N.~G., Carlin, B.~P., and Van Der~Linde, A. (2002).
\newblock Bayesian measures of model complexity and fit.
\newblock {\em Journal of the Royal Statistical Society: Series B (Statistical
  Methodology)}, 64(4):583--639.

\bibitem[{Stan Development Team}, 2015]{Stan2015}
{Stan Development Team} (2015).
\newblock Stan: A {C++} library for probability and sampling, version 2.7.0.

\bibitem[Staub, 2010]{staub2010eye}
Staub, A. (2010).
\newblock Eye movements and processing difficulty in object relative clauses.
\newblock {\em Cognition}, 116(1):71--86.

\bibitem[Traxler, 2014]{Traxler2014}
Traxler, M.~J. (2014).
\newblock Trends in syntactic parsing: anticipation, bayesian estimation, and
  good-enough parsing.
\newblock {\em Trends in cognitive sciences}, 18(11):605--611.

\bibitem[Traxler et~al., 2002]{traxler2002processing}
Traxler, M.~J., Morris, R.~K., and Seely, R.~E. (2002).
\newblock Processing subject and object relative clauses: Evidence from eye
  movements.
\newblock {\em Journal of Memory and Language}, 47(1).

\bibitem[Vandekerckhove et~al., 2011]{VandekerckhoveTuerlinckx2011}
Vandekerckhove, J., Tuerlinckx, F., and Lee, M.~D. (2011).
\newblock Hierarchical diffusion models for two-choice response times.
\newblock {\em Psychological methods}, 16(1):44.

\bibitem[Vasishth, 2015]{Vasishth:MScStatistics}
Vasishth, S. (2015).
\newblock A meta-analysis of relative clause processing in {M}andarin {C}hinese
  using bias modelling.

\bibitem[Vasishth et~al., 2013]{VasishthetalPLoSOne2013}
Vasishth, S., Chen, Z., Li, Q., and Guo, G. (2013).
\newblock Processing {C}hinese relative clauses: {E}vidence for the
  subject-relative advantage.
\newblock {\em PLoS ONE}, 8(10):1--14.

\bibitem[Vehtari and Gelman, 2015]{VehtariGelman2015Pareto}
Vehtari, A. and Gelman, A. (2015).
\newblock Pareto smoothed importance sampling.
\newblock {\em arXiv preprint arXiv:1507.02646}.

\bibitem[Vehtari et~al., 2015a]{VehtariEtAl2015Loo}
Vehtari, A., Gelman, A., and Gabry, J. (2015a).
\newblock Efficient implementation of leave-one-out cross-validation and waic
  for evaluating fitted bayesian models.
\newblock {\em arXiv preprint arXiv:1507.04544v2}.

\bibitem[Vehtari et~al., 2015b]{loo}
Vehtari, A., Gelman, A., and Gabry, J. (2015b).
\newblock {\em loo: Efficient leave-one-out cross-validation and {WAIC} for
  {B}ayesian models}.
\newblock R package version 0.1.3.

\bibitem[Vehtari and Ojanen, 2012]{VehtariOjanen2012}
Vehtari, A. and Ojanen, J. (2012).
\newblock A survey of {Bayes}ian predictive methods for model assessment,
  selection and comparison.
\newblock {\em Statist. Surv.}, 6(0):142–228.

\bibitem[Wagenmakers et~al., 2010]{WagenmakersEtAl2010}
Wagenmakers, E.-J., Lodewyckx, T., Kuriyal, H., and Grasman, R. (2010).
\newblock {Bayes}ian hypothesis testing for psychologists: A tutorial on the
  savage–dickey method.
\newblock {\em Cognitive Psychology}, 60(3):158–189.

\bibitem[Wang and Gelman, 2014]{WangGelman2014difficulty}
Wang, W. and Gelman, A. (2014).
\newblock Difficulty of selecting among multilevel models using predictive
  accuracy.
\newblock {\em Statistics at its Interface}, 7:1--8.

\bibitem[Watanabe, 2009]{Watanabe2009}
Watanabe, S. (2009).
\newblock {\em Algebraic geometry and statistical learning theory}, volume~25.
\newblock Cambridge University Press.

\bibitem[Watanabe, 2010]{Watanabe2010}
Watanabe, S. (2010).
\newblock Asymptotic equivalence of {Bayes} cross validation and widely
  applicable information criterion in singular learning theory.
\newblock {\em The Journal of Machine Learning Research}, 11:3571--3594.

\bibitem[Xu and Tenenbaum, 2007]{XuTenenbaum2007}
Xu, F. and Tenenbaum, J.~B. (2007).
\newblock Sensitivity to sampling in {B}ayesian word learning.
\newblock {\em Developmental science}, 10(3):288--297.

\end{thebibliography}

\end{document}